\documentclass{vgtc}                     

\usepackage{amsmath,amsfonts}
\usepackage{algorithmic}
\usepackage{algorithm}
\usepackage{array}
\usepackage{textcomp}
\usepackage{stfloats}
\usepackage{url}
\usepackage{verbatim}
\usepackage{graphicx}
\usepackage[export]{adjustbox}
\usepackage[skins]{tcolorbox}
\usepackage{cite}
\usepackage{xcolor}
\usepackage{tabu}                      
\usepackage{booktabs}                  
\usepackage{lipsum}                    
\usepackage{mwe}                       

\usepackage{mathptmx}

\onlineid{1909}

\vgtccategory{Research}

\title{HCVR Scene Generation: High Compatibility Virtual Reality Environment Generation for Extended Redirected Walking}

\author{%
  Yiran Zhang,
  Xingpeng Sun, and 
  Aniket Bera
}

\authorfooter{
  \item
  	Yiran Zhang is with Purdue University.
  	E-mail: zhan5058@purdue.edu
  \item
  	Xingpeng Sun is with Purdue University.
  	E-mail: sun1223@purdue.edu.

  \item Aniket Bera is with Purdue University.
  	E-mail: aniketbera@purdue.edu.
}

\abstract{%

Natural walking enhances immersion in virtual environments (VEs), but physical space limitations and obstacles hinder exploration, especially in large virtual scenes. Redirected Walking (RDW) techniques mitigate this by subtly manipulating the virtual camera to guide users away from physical collisions within pre-defined VEs. However, RDW efficacy diminishes significantly when substantial geometric divergence exists between the physical and virtual environments, leading to unavoidable collisions. Existing scene generation methods primarily focus on object relationships or layout aesthetics, often neglecting the crucial aspect of physical compatibility required for effective RDW. To address this, we introduce HCVR (High Compatibility Virtual Reality Environment Generation), a novel framework that generates virtual scenes inherently optimized for alignment-based RDW controllers. HCVR first employs ENI++, a novel, boundary-sensitive metric to evaluate the incompatibility between physical and virtual spaces by comparing rotation-sensitive visibility polygons. Guided by the ENI++ compatibility map and user prompts, HCVR utilizes a Large Language Model (LLM) for context-aware 3D asset retrieval and initial layout generation. The framework then strategically adjusts object selection, scaling, and placement to maximize coverage of virtually incompatible regions, effectively guiding users towards RDW-feasible paths. User studies evaluating physical collisions and layout quality demonstrate HCVR's effectiveness with HCVR-generated scenes, resulting in 22.78 times fewer physical collisions and received 35.89\% less on ENI++ score compared to LLM-based generation with RDW, while also receiving 12.5\% higher scores on user feedback to layout design.

}


\teaser{
  \centering
  \includegraphics[width=\linewidth]{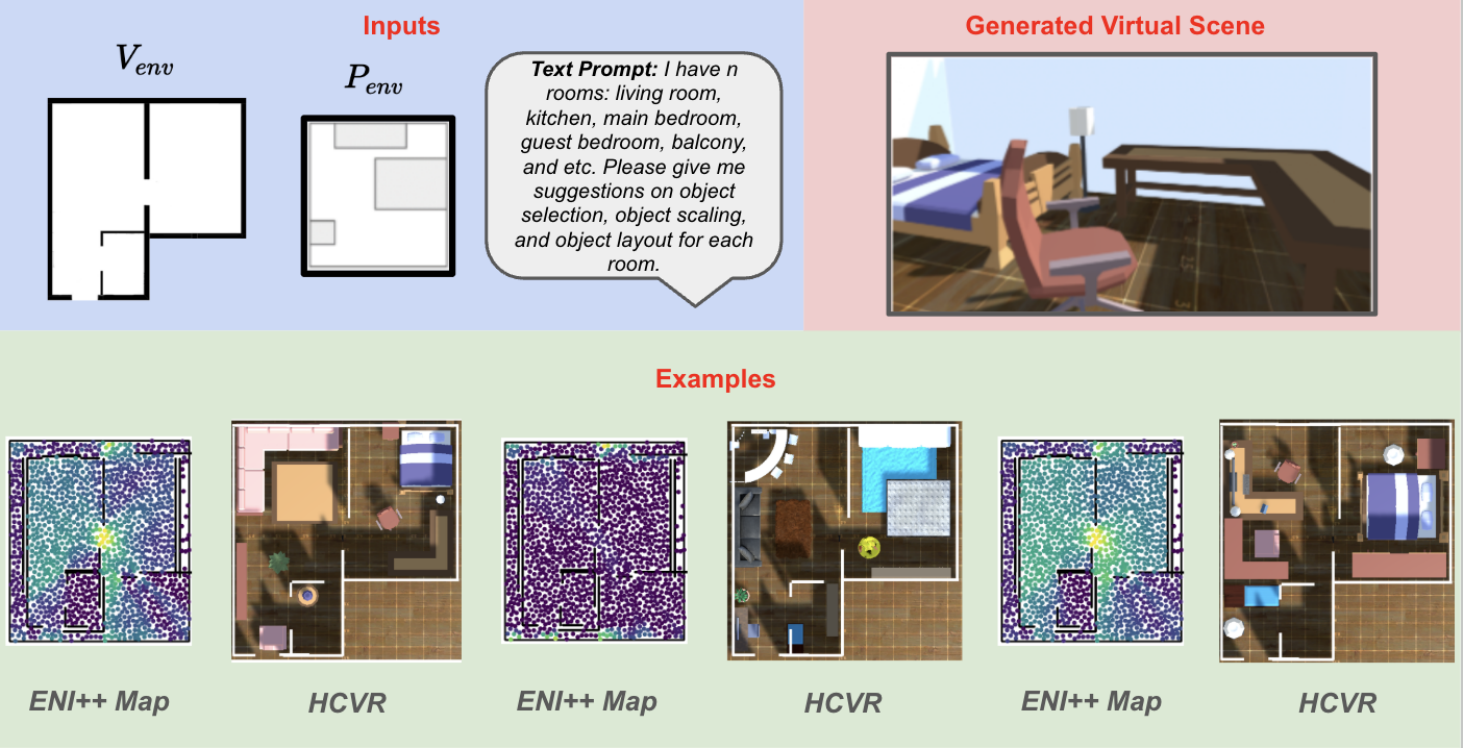}
  \caption{This is an overview of our HCVR virtual scene generation framework optimized for VR navigation and compatible with alignment-based redirected walking controller to provide user a unique natural walking interface. Given a virtual $V_{env}$ and physical $P_{env}$ scene pair and a text prompt, our model processes ENI++ score map to evaluate the scene compatibility in each area. The ENI++ scene plot shows sampled points colored from yellow (high ENI++ score as not redirected walking friendly) to purple (low ENI++ score as more redirected walking friendly). The blue block indicates the input to HCVR architecture while the red block shows an example of the generated virtual scene. The green block presents some more HCVR directed generated based on different ENI++ score plot.} 
  \label{fig:teaser}
}

\graphicspath{{figs/}{figures/}{pictures/}{images/}{./}} 

\usepackage{tabu}                      
\usepackage{booktabs}                  
\usepackage{lipsum}                    
\usepackage{mwe}                       

\usepackage{mathptmx}                  

\begin{document}


\firstsection{Introduction}

\maketitle

Natural walking in virtual environments offers a unique exploration experience for VR users \cite{usoh1999walking, bruder2012redirecting}. It enables users to engage with virtual environments naturally to improve their task performance \cite{ruddle2009benefits}. Previous research introduced the redirected walking (RDW) techniques and indicates its ability to provide users a free exploration of virtual environments within constraint physical spaces \cite{williams2021arc,williams2021redirected,thomas2019general,hodgson2013comparing}. RDW algorithms adjust the rotation and translation of the virtual scene, redirecting users onto collision-free paths without being noticed. These RDW approaches, such as steering toward the center \cite{razzaque2002redirected}, orbiting or targeting objects \cite{razzaque2005redirected}, potential field-based redirection \cite{thomas2019general}, reinforcement learning-based redirection \cite{shibayama2020reinforcement, lee2019real}, and alignment-based redirection \cite{williams2021arc}, have been proven to be effective in various scenarios. Nevertheless, the extent of redirection in translation and rotation is limited by human perception threshold \cite{jung2019redirected, schmitz2018you, grechkin2016revisiting, williams2019estimation, hutton2018individualized} to prevent users from noticing the redirection during the virtual experience \cite{hildebrandt2018get}. This thus lead to unpreventable collisions while users' physical space is significantly incompatible with a designed virtual space.

Previous research has established that the geometric structure and shape of an environment significantly influence user navigation behavior \cite{simpson2017quantifying, dalton2010judgments, franz2005exploring}. In addition, studies have explored the feasibility of collision-free navigation based on distinct virtual environment layouts \cite{wiener2007isovist}. Therefore, instead of directly applying RDW controller upon existing virtual scene, we suggests to generate virtual scenes that is compatible with users' physical surrounding and fully release the advantage of RDW controller to lead user onto a safe virtual walking experience. 

Existing 3D scene generation approaches primarily focus on visually appearing and spatially reasonable. Some generated scenes are either modified from pre-defined designs \cite{wang2021sceneformer,yang2024scenecraft} or trained to learn the spatial relationships between objects \cite{yang2022modeling,yang2021layouttransformer}. However, these methods are constrained by the extensive effort required to manually track spatial relationships across a wide range of object categories or obtain high requirement for the diversity of the layouts that dataset contains. Leveraging the strong contextual understanding capabilities of Large Language Models (LLMs), recent LLM-based scene generation methods \cite{yang2024holodeck} significantly alleviate the burden of explicitly specifying objects' spatial relationships, while enabling greater diversity in scene configurations without the necessary of the layout labeling or training process. Despite these advances, existing scene generation methods are not specifically designed to support natural walking interfaces in virtual reality. To address this gap, we propose the HCVR architecture, which generates virtual scenes that are both spatially coherent and optimized for redirected walking (RDW) applications. We also demonstrate the advantages of HCVR-generated scenes in terms of layout quality and their ability to facilitate collision-free redirected walking paths.

Figure \ref{fig:architecture} presents an overview of the HCVR architecture. Given a user's physical environment, HCVR assesses the redirected walking feasibility of virtual floor plans using a boundary-sensitive and interpretable Environment Navigation Incompatibility metric (ENI++), an extension of the original ENI metric\cite{williams2022eni}. After generating an ENI++ score map to visualize the compatibility of different regions in the virtual space, HCVR employs BUDAS \cite{gan2021many} to segment rooms within the virtual floor plan. Subsequently, HCVR utilizes LLM to select appropriate 3D assets, defines relational constraints among selected objects, and scales the assets in a visually coherent and spatially consistent manner. Moreover, HCVR incorporates an Out-of-Boundary (OOB) checker to ensure that all placed objects remain within their corresponding virtual room boundaries. A final adjustment phase further refines object positioning and scaling to maximize the availability of redirected-friendly virtual areas, thereby enhancing the potential for effective redirected walking.

In summary, our main contributions are as follows:
\begin{itemize}
\item A novel interactive VR environment generation framework (HCVR) that utilizes ENI++ for LLM-guided scene generation so that the resulting virtual scene is reasonably designed and redirected friendly.
\item The ENI++ metric extends the original ENI \cite{williams2022eni} by incorporating virtual boundaries' and obstacle borders' constraints, and rotation-sensitive visibility polygon comparisons, thereby providing a more describable and effective measure of compatibility between physical and virtual spaces. 
\item Extensive user study experiments (N = 16) on the satisfaction of ENI++ -generated VR scenes and collision-free navigation in the output scenes. HCVR-directed scenes outperform the baseline LLM-generated scene by 7.75\%  on the overall layout feedback. It also receives the same level of feedback on object selection and 3.38\% better on object scaling. Although it sacrifices 3.72\% satisfaction in positioning and rotation, redirecting users in ENI++ -generated scene avoids 22.78 times more physical collisions than redirecting users in LLM-generated scene across five distinctly designed experiments.
\end{itemize}

\section{Background \& Related Works}
\subsection{Virtual Reality Scene Generation}
\subsubsection{Scene Layout Generation}
Scene layout generation has evolved significantly with the integration of deep learning techniques, particularly focusing on structured inputs like scene graphs and label sets. Early approaches such as LayoutVAE employed variational autoencoders to model object layouts stochastically from label sets, enabling probabilistic generation \cite{jyothi2019layoutvae}. Subsequent advancements like LayoutTransformer \cite{gupta2021layouttransformer} leveraged self-attention mechanisms to capture contextual relationships between layout elements across diverse domains, supporting both generation and completion tasks. Recent works emphasize conceptual and spatial diversity: LT-Net\cite{yang2021layouttransformer} introduced transformer-based architectures to infer implicit object relationships from scene graphs, while novel-view prediction models\cite{qiao2022learning} explored layout synthesis across camera viewpoints using object context transformation modules. For 3D scenes, SceneCraft demonstrated layout-guided generation through multi-view proxy maps and diffusion models, enabling complex indoor environments beyond single-room constraints \cite{yang2024scenecraft}. These developments highlight a trend toward combining geometric reasoning with semantic understanding to address both 2D and 3D layout synthesis challenges. Nevertheless, these approaches remain constrained by the need to predefined spatial relationships or the limited spatial diversity present in the training data.

\subsubsection{LLM Layout Generation Methods}
Recently, Large Language Models (LLMs) have demonstrated their effectiveness in context reasoning across various fields \cite{sun2024trustnavgpt, sunbeyond, pham2024mvgaussian, guhan2024tame}. Recent studies have also used LLM for creating room layouts \cite{feng2024layoutgpt, lin2023towards, ccelen2024design, yang2024holodeck}. However, limitations in the understanding of LLMs of numerical values \cite{guo2024learning, zhang2024mathverse} and spatial relationships \cite{yamada2023evaluating} can lead to positional issues, such as object overlaps and objects out-of-boundary (OOB) problems \cite{feng2024layoutgpt, lin2023towards}.

Similarly, both Holodeck \cite{yang2024holodeck} and I-Design \cite{ccelen2024design} incorporate spatial constraints in their layout prompts and apply optimization techniques to address out-of-boundary (OOB) issues after generating the layout. I-Design employs eight spatial relations:left, right, front, behind, on, above, under, and corner to describe object placement, while Holodeck uses 10 spatial relations across five categories: global, distance, position, alignment, and rotation. Both systems implement algorithms to adjust object positions as a post-processing step to correct spatial errors generated by GPT-based layout methods, demonstrating effectiveness in producing reasonable layouts and resolving spatial issues. However, these methods are not specifically designed for virtual layout generation and fail to account for natural walking compliance in VR setups or the spatial discrepancies between the virtual and physical environments.

\subsubsection{Scene Modification for VR Navigation}
Tailored Reality \cite{dong2021tailored} presents an innovative algorithm to adjust virtual environments in order to align with a user's physical surroundings. Traditional virtual reality (VR) scenes are typically pre-designed, often leading to discrepancies when users attempt to navigate under physical situation. This misalignment can disrupt the immersive experience and hinder natural movement.
To address this challenge, the authors introduce a perception-aware retargeting technique that scales and slightly re-positions objects in virtual scenes for user to explore without interruptions or visual distortions. The algorithm takes into account various factors, including dynamic visibility and object relationships from a first-person perspective, ensuring that the virtual space remains coherent and navigable.

This method still need a pre-designed virtual scene for modification process and modification on existing incompatible virtual scenes still limits redirected walking controllers' ability. To overcome these limitations, our architecture HVCR incorporates the ENI++ metric alongside LLM-based layout suggestions. This integration minimizes the average ENI++ score across unoccupied virtual regions designated for user navigation, thereby facilitating more effective redirection along collision-free paths. Human evaluation studies also demonstrate high user satisfaction with the resulting LLM-suggested layouts, underscoring the effectiveness of our method.

\begin{figure*}[thpb]
\centering
\vskip -0.2in
\centerline{\includegraphics[width=\textwidth]{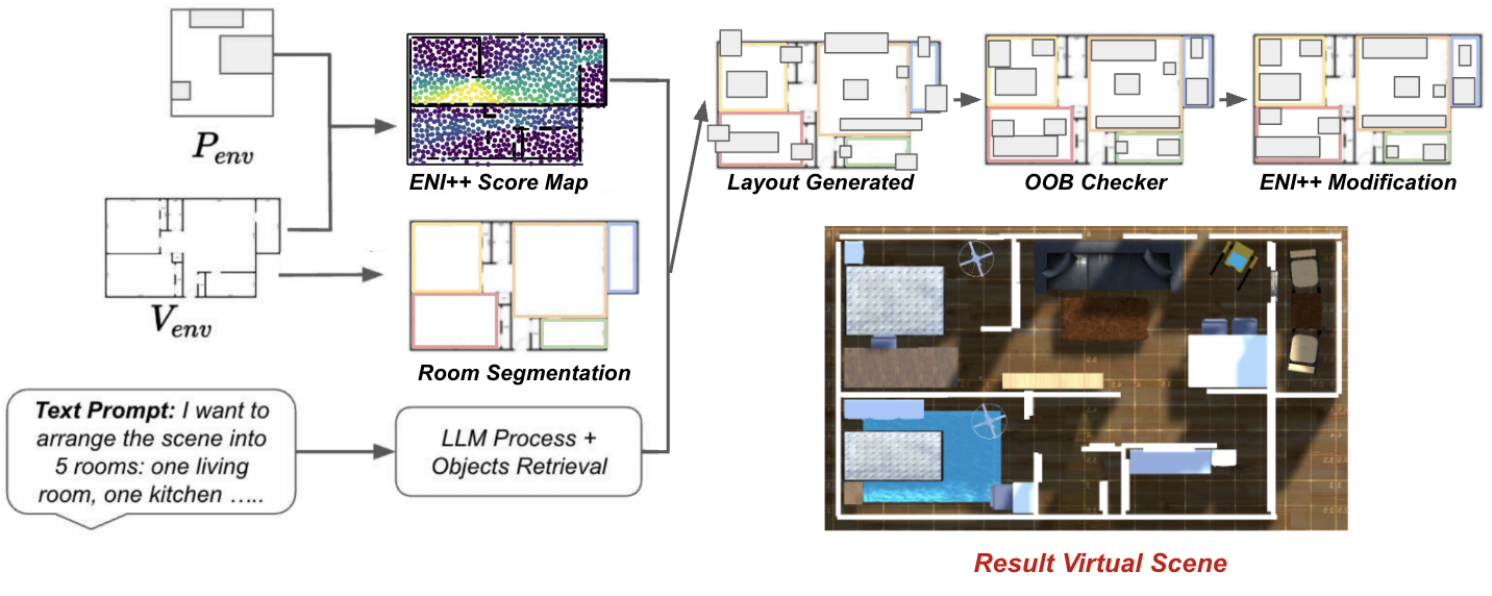}}
\vskip -0.1in
\caption{\textbf{HCVR Architecture}: A physical environment with obstacles $P_{env}$ and a virtual space with a floor plan $V_{env}$ is able to use ENI++ metric to process a score map. We use BUDAS \cite{gan2021many} to identify the rooms in the virtual floor plan. The objects for each room,  their spatial relationships between each others, and the room will thus be generated. The compatibility information for each room processes along with the LLM prompt in the initial layout generation. Sampled points are displayed across the scene plot, with colors transitioning from yellow to purple to represent ENI++ scores—yellow denoting high values and purple indicating low ones. Out-Of-Boundary (OOB) checker thus keep all objects in the initial layout inside each room without any overlaps. Lastly, we retrieve the score map to re-modify the objects locations within their spatial constraints so that all of them can cover as much incompatible area as possible. HCVR thus finally left the compatible and redirected friendly area available for user to explore.}
\label{fig:architecture}
\vskip -0.2in
\end{figure*}

\subsection{Natural Walking in Virtual Scene}

Natural walking provides an intuitive means of exploring virtual spaces, prompting researchers to develop techniques that extend the available virtual space beyond the user's limited physical environment. One such area of study focuses on visual-based redirection methods. Sun et al. \cite{sun2018towards} introduced advances in continuous natural movement within VR environments through saccadic redirection. Their work focuses on dynamic saccadic redirection, a novel approach that utilizes users' rapid eye movements (saccades) to imperceptibly alter their walking paths, enabling an infinite walking experience within finite physical spaces. Based on this, Joshi et al. \cite{joshi2023enabling} optimized saccadic redirection strategies, reducing latency and making the redirection process more seamless. In addition, researchers have explored the use of visual distractors to enhance redirected walking (RDW) experiences in VR \cite{cools2019investigating,peck2010improved,chen2017supporting}. These distractors also contribute to the naturalness and immersion of the virtual walking experience, supporting more effective RDW implementations.

Perception threshold gains refer to the allowable degree of mismatch between the physical and virtual walking of users in terms of rotation, speed, and curvature \cite{steinicke2008analyses, steinicke2009estimation, bruder2012redirecting, matsumoto2020detection, grechkin2016revisiting, kruse2018can, paludan2016disguising}. These thresholds have led to the development of various redirected walking (RDW) algorithms \cite{razzaque2005redirected, thomas2019general, shibayama2020reinforcement, lee2019real, williams2021arc}, which update the location and rotation of the head-mounted display (HMD) in the virtual space per frame.

Steering methods in RDW, such as steering toward the center, orbiting and targeting, redirect users either around the center of the physical space, along a circular path around the physical space, or toward a target physical location \cite{razzaque2005redirected}. Potential field-based redirection \cite{thomas2019general} introduces a method to steer users away from physical boundaries and obstacles by constructing an artificial potential field within physical space. Reinforcement learning has also been used to optimize redirected paths, using collision detection training simulations to refine redirection strategies \cite{shibayama2020reinforcement, lee2019real}. Another RDW algorithm \cite{williams2021arc} focuses on aligning the physical distance of the users from obstacles with their virtual distance of objects in the direction they are facing, further enhancing the realism of the virtual navigation experience. 

All RDW techniques are designed to mitigate collisions within a predefined virtual and physical environment configuration. The existing analysis of these algorithms often overlooks the influence of the compatibility of the physical environment on collisions within the virtual space. In this study, we conducted a comparative evaluation between a low-compatibility virtual environment design and a high-compatibility design generated using the ENI++ framework, both employing the same RDW algorithm. Furthermore, our results highlight that, under some experiments, the compatibility of the virtual environment design can exert a more significant impact on collision outcomes than the choice of the RDW algorithm itself.


\subsection{Virtual Environment Compatibility}
Environment Complexity metric are used to measure the ability of user to perform a task in virtual environments on different topics. ENI \cite{williams2022eni} offers a method to quantify the ability of a user to walk freely in virtual spaces. The shape matching of both visibility polygons in physical and virtual spaces can describe the ability to walk freely under redirection at the current position and rotation of the user \cite{williams2021redirected}. ENI calculates all these shape matchings of sampled points in each pair of physical and virtual environment. An ENI score map can therefore represent the possibility of free-walking in each area in a virtual space. ENI++ can be considered as an updated version of ENI. Instead of considering the overall map, ENI++ only calculates the visibility polygons in reachable displacement. ENI++ also limit the rotation calculation to be boundary sensitive.

Environment Complexity metric are used to assess users' ability to perform tasks in virtual environments across various contexts. The ENI metric \cite{williams2022eni} provides a method to quantify a user’s ability to navigate freely in virtual spaces. By comparing the shape matching of visibility polygons in both physical and virtual spaces, ENI assesses the feasibility of free walking under redirection in a user’s current position and rotation \cite{williams2021redirected}. Calculate these shape matchings for sampled points across pairs of physical and virtual environments, generating an ENI score map that represents the possibility of free walking in different areas of a virtual space.

ENI++ metric is similar to the ENI metric, as both generate score maps for pairs of physical and virtual environments to represent the compatibility of each virtual area with the corresponding physical space. Instead of evaluating the entire environment, ENI++ focuses on the reachable empty space within a limited distance and restricts the rotation calculation to be boundary sensitive, offering a more localized and refined measure of the free-walking potential.


\section{Methodology}
\subsection{Overview}
Given a physical environment configuration $P_{env}$, a virtual floor plan $V_{env}$, and a textual scene description, our method generates a virtual layout optimized for physical walking(Figure \ref{fig:architecture}). We detail: (1) the ENI++ metric in \ref{sec:ENI++}, (2) LLM-based object and relationship generation in \ref{sec:llmlayout}, (3) the Out-of-Boundary (OOB) checker in \ref{sec:oob}, and (4) final object adjustment using the ENI++ score map in \ref{sec:final}.

\subsection{ENI++: An Extension of ENI Metric}\label{sec:ENI++}

Our Environment Navigation metric ENI++ is expanded from ENI \cite{williams2022eni} metric that quantifies the physical compatibility of a virtual region. Similar to ENI, we sample points from both environments such that $V_p \in V_{env}$ and $P_p \in P_{env}$ for a given virtual and physical environment pair $(V_{env},P_{env})$. For each sampled point pair $(v_p, p_p)$, where $v_p \in V_p$ and $p_p \in P_p$, we define $v_{poly}$ and $p_{poly}$ as the polygonal areas constrained by translation distance from $v_p$ and $p_p$. Each polygon pair ($v_{poly}$,$p_{poly}$) compares to each other with reachable distances and rotations to maximize the overlapped area. The ENI++ score for each sampled point $v_p$ is calculated as the minimum of all the maximum similarity observed area among $P_p$.

\begin{figure}[thpb]
\vskip -0.1in
\centering
\centerline{\includegraphics[width=\columnwidth]{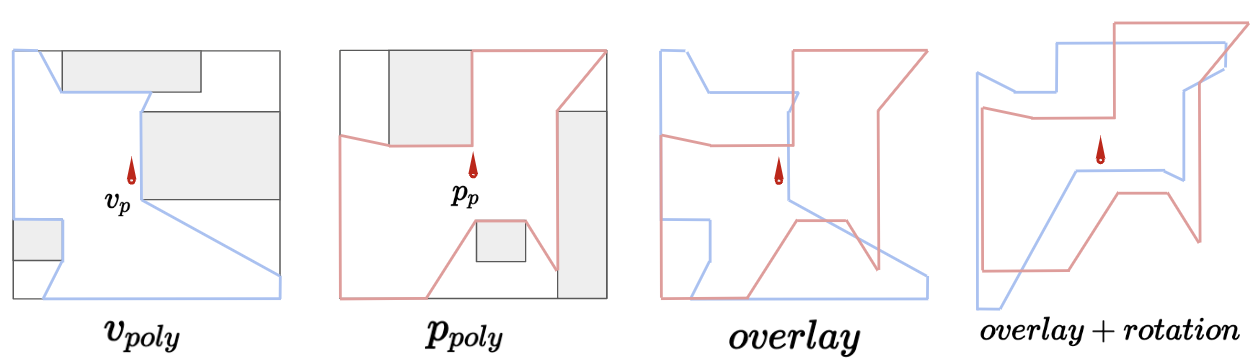}}
\caption{The blue bounded area indicates a calculated visibility polygon from point $v_p$ and the red bounded area represents one from $p_p$. The $overlay$ area of both visibility polygons is constrained and not allowed the user to walk into the current direction due to the front obstacle in $P_{env}$. However, this problem can be solved if the user is being redirected and walking with a certain degree of rotation. The $overlay + rotation$ clearly enlarged the overlapped area and allows a larger accessible area for users.}
\label{fig:vispos}
\vskip -0.1in
\end{figure}

\subsubsection{Points selection}

The available area of both the $V_{env}$ and $P_{env}$ environments is bounded by the given virtual space boundary and the obstacles inside. In our test environments, the initial input $V_{env}$ is floor plans, and thus the available area is constrained by the outline of the given floor plan and the wall obstacles detected inside. The boundary and obstacles in $P_{env}$ should align with the physical environments of each VR user. Instead of random sampling, we used Delaunay triangulation \cite{chew1987constrained} to ensure that points are evenly sampled in the non-obstacle area of both environments. We restricted the size of each triangle in the Delaunay triangulation process based on the size of each environment so that the sampling result can be valid for different $(V_{env},P_{env})$ pairs. The points produced during the Delaunay triangulation are therefore selected as $V_p$ and $P_p$ for $V_{env}$ and $P_{env}$.

\subsubsection{Visibility Polygons and Similarity Comparison}
 In ENI++, we only consider the reachable area comparison, which means it only compares the virtual visibility polygons under available virtual translations and rotations with the corresponding human threshold available physical visibility polygons. Comparison of local visibility polygons can lead to a more describable result when a large or complex $V_{env}$ is given. 

For any given point $v_p$ or $p_p$ in an environment, the visibility polygon $v_{poly}$ or $p_{poly}$ is an area on the plane that is always visible from $v_p$ or $p_p$ as Figure \ref{fig:vispos} shows.

\begin{figure}[thpb]
\centering
\centerline{\includegraphics[width=0.45\textwidth]{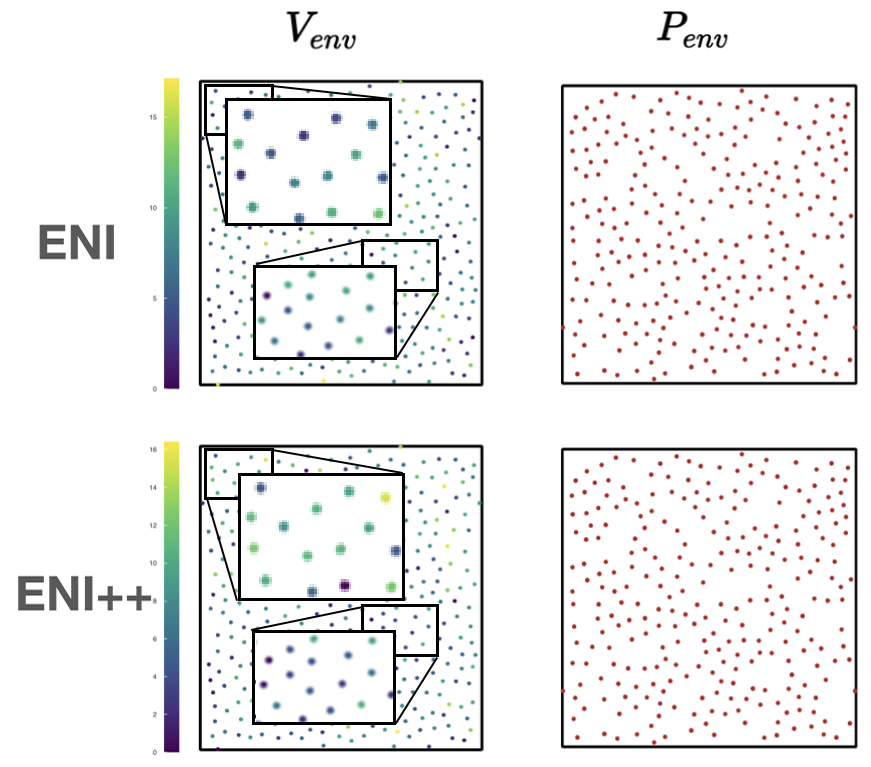}}
\caption{This figure illustrates the differences between ENI and ENI++ in the resulting plots with an identical virtual and physical scene pair. The virtual environment $V_{env}$ is defined as a 20×20 square, while $P_{env}$, is a 20x25 rectangular area. The zoomed in patched shows the difference between ENI and ENI++ due to the implementation of boundary-sensitive and rotation-sensitive visibility polygons comparison.}
\label{fig:eniENI++}
\end{figure}

We interpolate all available rotations at each $v_p$, and use rotation gains to keep the intensity of physical redirection within the perception threshold. For each virtual rotational direction $v_{rot}$, there are corresponding possible redirected physical rotations $p_{rot}$ in the range of the minimum rotation gain $RG_{min}$ at 0.67 and the maximum rotation $RG_{max}$ at 1.24 \cite{williams2019estimation}. A setting of 4 meters * 4 meters is also applied as the default local translation distance in our experiments in \ref{sec:experiments} due to the consideration of computational efficiency. The visibility polygon that constructed at each $v_p$ can thus be compared with each $p_p$ under the rotation of $(RG_{min},RG_{max})$, which scales the virtual rotational direction. For each $v_p$, we interpolated 36 directions out of 360 degrees as $v_{rot}$s. The visibility polygon of a pair $(v_p,p_p)$ only finds a shape similarity between all 36 directions of $v_{poly}$ with each direction's 10 rotations within rotation gain $(RG_{min},RG_{max})$ and translation of $p_{poly}$. Shape similarity $v_{poly}$ and $p_{poly}$ is the overlap between two areas and therefore can be calculated as \cite{williams2021redirected}:
\begin{equation}
    S(v_{poly},p_{poly}) = v_{poly} \setminus (v_{poly} \setminus p_{poly})
\end{equation}

\begin{figure}[h]
\centering
\centerline{\includegraphics[width=\columnwidth]{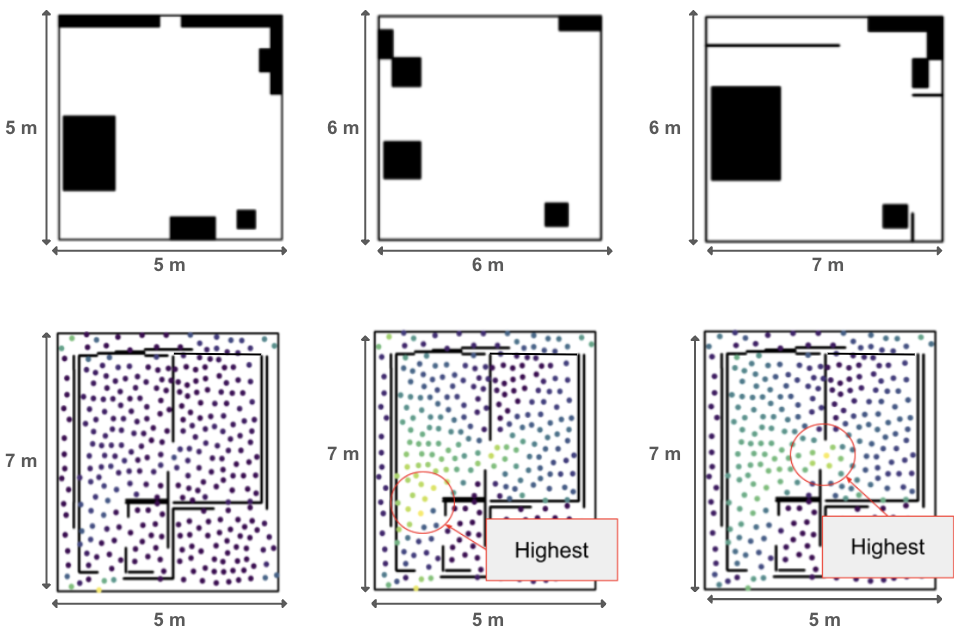}}
\caption{This figure shows different ENI++ score map being plot on the same $V_{env}$ while given distinct $P_{env}$. The ENI++ score was plotted as colors ranges from purple to yellow (low to high). It is obvious to see how it varies. The first scene is shows high compatibility. The second pair shows the lowest ENI++ score as the marked area at the left while the third ENI++ plot shows the lowest score in the middle area.}
\label{fig:ENI++Score}
\vskip -0.2in
\end{figure}

The ENI++ score of one $v_p$ with one $p_p$ can be thus calculated as Algorithm \ref{alg:ENIAlgo}. After applying the Algorithm \ref{alg:ENIAlgo} with all the $v_p$s with all $p_p$s, we are able obtain a overall ENI++ score for all the $p_p$s: 
\begin{equation}
    \text{ENI++}(v_p) = \arg\min \{ \forall p_{p} \in P_{p},\; \text{ENI++}(v_p,p_{p}) \}.
\end{equation}
ENI++ is boundary and corner inclusive for virtual space's borders and obstacles. For each $v_p$ around the virtual boundary border line and the obstacle border line, the virtual rotation towards these border lines will be ignored. Then, ENI++ score maps can be obtained as Figure \ref{fig:ENI++Score} shows. This figure also illustrates the compatibility of the same $V_{env}$ under different $P_{env}$ s and obstacles installed inside.

\begin{algorithm}
\caption{ENI++ Score of $v_p,p_p$}
\label{alg:ENIAlgo}
\begin{algorithmic}[1]
\REQUIRE Compute $v_{\text{poly}}$ for $v_p$ and $p_{\text{poly}}$ for $p_p$.
\FORALL{Head directions $v_{\text{rot}}$ from $0^\circ$ to $360^\circ$}
    \STATE Initialize $ShapeCompare \gets [\,]$
    \FORALL{Rotations $p_{\text{rot}}$ from $RG_{\min}$ to $RG_{\max}$}
        \STATE $ShapeCompare.\text{add}(S(v_{\text{poly}}, p_{\text{rot}}(p_{\text{poly}})))$
    \ENDFOR
    \STATE ENI++$(v_p, v_{\text{rot}}) \gets \arg\max \{ShapeCompare\}$
\ENDFOR
\STATE ENI++$(v_p, p_p) \gets \text{avg}\{ \forall v_{\text{rot}}, \; \text{ENI++}(v_p, v_{\text{rot}}) \}$
\RETURN{} 
\end{algorithmic}
\end{algorithm}

Since ENI++ is sensitive to boundaries, score differences near the edges are more pronounced. In contrast, variations in the central region are expected to be minimal (Figure \ref{fig:eniENI++}). This is because ENI computes the similarity by evaluating all possible rotations of each virtual point, which yields results generally aligned with those from ENI++, as both ultimately account for shape similarity across rotations. Moreover, ENI++ introduces an additional constraint on rotational gain, allowing it to generate more precise and informative plots. For example, Figure \ref{fig:eniENI++} shows the ENI++ patches shows a middle of the virtual space more compatible (with darker color) whereas the corner of the virtual space is less compatible (with lighter color). The ENI++ generated a more accurate map because the corner is harder to be redirected under some $v_rot$s and the middle of an empty space is more redirected friendly.

\subsection{LLM Layout Generation}\label{sec:llmlayout}
The input virtual scene consists of walls and borders as a floor plan. A BUDAS room detection algorithm \cite{gan2021many} can output a room segmentation result that includes the number of rooms, room areas, and positions. Using the ENI++ metric, we can calculate an initial capacity score $R_{score}$ of each room $R$ as below:
\begin{equation}
    R_{score} = Avg(\sum^{v_p \in R} \text{ENI++}(v_p)).
\end{equation}
Each room capacity score $R_{score}$ is based on the average ENI++ scores of all $v_p$s sampled in the room. We then normalize this capacity score in the range of 1 to 10 where a larger number representing more area need to be occupied. The size of the room $m$ calculates as $\dfrac{Area(R)}{Area(Virtual)}$. Since most LLM model is more sensitive to context language instead of number, the scale of a room can thus be separated into five scales: huge ($m>0.25$), large ($0.166<m<0.25$), medium ($0.125<m<0.166$), small ($0.1<m<0.125$), and tiny ($m<0.1$). The description of room sizes and their capacity scores are variable inputs into LLM prompt to aid with later object selection and scaling procedure.

The LLM text prompt aids in selecting 3D assets for a virtual scene, positioning all the assets reasonably, and scaling the assets in a visually sensible way. Given an LLM text prompt $T$ that includes capacity scores, size description, and functions of all rooms (Figure \ref{sec:prompt_first}), the LLM is able to select reasonable size and number of objects from assets so that it can work for covering all the high ENI++ score areas. In LLM object selection procedure, the function of a room helps generate a list of object names. The capacity score determines how many items in minimum a room should have. In experiments, we utilize 
\begin{equation}
    R_{score}/({\dfrac{Area(Virtual)}{Area(Physical)}}) 
\end{equation}
as a minimum number of items for each room. Then, we asked the LLM to find the best fit object names in 3D Model assets as selected objects for each room. We use in-context training in the LLM to ensure it chooses objects that have compatible sizes with the room (Figure \ref{sec:incontext}). The size description also filters some large objects out during the asset selection. For instance, if a room is tiny, the length X width of the selected objects cannot take more than one tenth area of the scene. 

\begin{figure}[!ht]
\begin{tcolorbox}[width=\linewidth]
\textbf{System Instruction:}\\
You are a skilled interior designer. Your task is to interpret the given instructions of each room, including the function and size description of a room, and the minimum number of items to put into a room. Your job is to name a list of objects that should appear in each room based on the instruction. Your answer must strictly follow the provided format. Do not write any additional text. \\
\textbf{Example:}\\
Input: An apartment contains: a huge living room with at least 5 items. 

Output: living room (sofa, dinner table, chairs, TV, bookshelf, painting decoration, coffee table). 
\end{tcolorbox}
\caption{Demo LLM prompt for object suggestion which uses capacity score as minimum number of items of a room. Function and size description of each room can impact the LLM output on items and number of items (complete prompt details are in supplementary materials).}
\label{sec:prompt_first}
\vskip -0.1in
\end{figure}

\begin{figure}[!ht]
\begin{tcolorbox}[width=\linewidth]
\textbf{In-Context Training}\\
\textbf{Example 1:}\\
Input: living room (sofa, dinner table, chairs, TV, bookshelf, painting decoration, coffee table). Room (3*3), sofa (1*4), dinner table(2*2), chairs(0.4*0.4), ...

Output: False, because the sofa one axis of sofa is significantly larger than the room. Change Suggestion: Sofa. Delete Suggestion: None.

\textbf{Example 2:}\\
Input: front door are (floor lamp, shoe rack, chair, clothing rack). Room (1*0.6), floor lamp (0.3*0.3), shoe rack (0.8*0.4), chair(0,5*0.5), clothing rack (0.3*1).

Output: False, because the total area is smaller than all the object combined. Change Suggestion: clothing rack,shoe rack. Delete Suggestion: clothing rack.
\end{tcolorbox}
\caption{Demo in-context training for object selection. This figure give examples of two cases: an object with one axis larger than the room, and total objects cover more area than the room area (more in-context training details included in supplementary materials).}
\label{sec:incontext}
\end{figure}

Then, with the selected items, we also query LLM to return reasonable spatial relationship between them (Figure \ref{sec:prompt_second}). The defined spatial relationship includes relationships between scenes (near/far to wall $X$, middle of the room, corner of wall $X$ and wall $Y$) and between objects (left, top, right, bottom, front, behind, top-left, top-right, bottom-left, bottom-right, top-center, and bottom-center of object $X$). The output of the given prompt includes a list of scaled objects for each room and their spatial relation between each other and the room. In all experiments showed in this paper, we use GPT-4o with default parameters.

\begin{figure}[!ht]
\begin{tcolorbox}[width=\linewidth]
\textbf{System Instruction:}\\
You are a skilled interior designer. Given a list of objects for each room, your job is to return your suggestion of their spatial relationship with each other and the environment. Your answer must strictly follow the provided format. Do not write any additional text. \\
\textbf{Example:}\\
Input: living room (sofa, dinner table, chairs, TV, bookshelf, painting decoration, coffee table).

Output: living room (sofa near upper wall, coffee table front of sofa, chair A left of dinner table, chair B right of dinner table, paining decoration near upper wall, bookshelf near to right wall, TV near to bottom wall).
\end{tcolorbox}
\caption{Demo LLM prompt for layout suggestion based on the given lists of objects for each room ( more prompt details are in supplementary materials).}
\label{sec:prompt_second}
\end{figure}

\subsection{Out-Of-Boundary Modification}\label{sec:oob}

\begin{figure}[thpb]
\centering
\centerline{\includegraphics[width=\columnwidth]{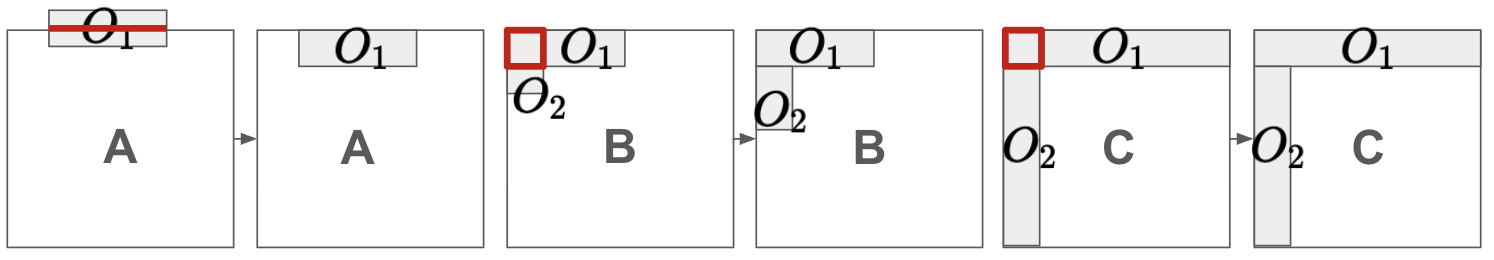}}
\caption{Step A: representing an object $O_1$ is positioned at the line and being moved within the boundary through modifying the y axis. Step B: presenting an detected overlap area (red square) and $O_2$ as a smaller one between the two overlapped objects being moved downwards. Step C: representing a case of unavoidable overlap between $O_1$ and $O_2$, $O_2$ as a larger object between the two was scaled down on the y axis only.}
\label{fig:OOB}
\vskip -0.2in
\end{figure}

After the generation of spatial relationships and an initial plot of the generation result, some objects can appear outside the boundary of the scene. For each room, the corresponding group of objects in this room first checks if there are assets outside the bound. If an object overlaps with the boundary or completely outside the boundary, we move that object along the overlapped edge or the edge close to it so that it can appear inside the room as Step A in Figure \ref{fig:OOB}. Then we check the overlapped objects inside each room and modify one axis at a time until they fit into the room without any overlaps, as Step B of Figure \ref{fig:OOB} shows. If an overlap of an object is inevitable, as in Step C of Figure \ref{fig:OOB}, we put them in positions that have the smallest overlap area and scale the overlapped axis down until they do not overlap. However, the scaling procedure is constrained to between (0.8, 1.2) of its original size in our experiments. If an overlap is still not avoidable, the model ignores the object and sends the object name back to the LLM for another object retrieval process.

\subsection{ENI++ Directed Room Modification}\label{sec:final}

After all objects appear in the expected room without any overlaps and OOB error, the ENI++ score map can apply a decent final gradient modification to ensure that the scene is physical walking friendly. Each object in the scene covers a certain number of sampled points $v_{covered} \in V_p$, and leaves others $(V_p \setminus v_{covered})$ as $v_{uncovered} \in V_p$. This minimizes the average of all points' ENI++ score that are unoccupied by objects in $V_{env}$, i.e.

\begin{equation}
   Min \{Avg \{ \text{ ENI++ }(V_p \setminus v_{covered})\}\}.
\end{equation}

  Meanwhile, we want assets to be allocated reasonably during the gradient descent process and thus constrained the modification based on the LLM spatial relationship output of each room where we check the spatial relationship between assets in each modifying iteration. After a decent amount of iterations, the objects placed in the virtual environment can remain highly physical environment compatible.

\section{Experiments}\label{sec:experiments}

\textbf{Participants }     We recruited 16 participants (ages 22–30, 8 female, 8 male) for the study. All had normal or corrected normal vision, no history of auditory impairments, and no current or former disabilities. Among them, 7 had no previous experience with VR, 5 had some VR experience, and 4 were familiar with and had regular access to VR technology. None were aware of the research hypothesis. The study was approved by the Institutional Review Board (IRB) of the host institution and all participants provided their informed consent before participation.

\textbf{Experimental Setup }     During the study, subjects wore a Meta Quest 2 HMD for visual display. Subjects expect to walk naturally in a large empty area to explore the virtual environment displayed in VR. During the test, user preference surveys were conducted and physical collisions with designed physical scenes were recorded in the system. Meanwhile, participants were asked to walk in a default scene for 1 minutes before the study to minimize the impact of their initial perception of the test area on their subsequent walking behavior.

During studies, users will not receive notifications upon colliding with the designed physical environment in the system to maintain a seamlessly natural walking experience. The number of physical collisions is recorded separately per user (=16), experiment (=5), and ablated study (=3).

\textbf{Experimental Conditions }     We conducted experiments in five pairs of physical and virtual environments. These five scenarios cover different sizes of rooms that include (1) a scene of $(P_{env} > V_{env})$, (2) a scene of $(P_{env} = V_{env})$, (3) a scene of $(P_{env} < V_{env})$ where the length of $V_{env}$ is half-time longer than that of $P_{env}$, (4) a scene of $(P_{env} < V_{env})$ where both the width and the length of $VE$ are half-time longer than that of $P_{env}$, and (5) a scene of $P_{env}$ where both the width and the length of $V_{env}$ are one time longer than that of $P_{env}$ (referring the experiment pair plot in appendix).

For each pair of $(P_{env}, V_{env})$, the ablation study between three generated cases was performed to validate the effectiveness of our layout design together with RDW can redirect users to a "collision-free" path. User satisfaction and collision data are collected across: (1) the LLM$\&$ARC: LLM-prompted scene + OOB checker along with the ARC \cite{williams2021arc} during scene exploration, (2) the ENI++: ENI++ metric guidance scene (including ENI++ score map input to LLM and ENI++ -directed modification procedures) without the RDW algorithm applied during user test, and (3) the ENI++$\&$ARC: ENI++ metric guidance scene with ARC \cite{williams2021arc} during scene exploration. Each subject in our user study thus had (\# scene pair = 5)×(\# condition = 3) = 15 trials. 

The five physical and virtual scene pairs are designed to test under various conditions. Experiment 1 describes a case where $(PE > VE)$. Experiment 2 indicates the physical collisions when $(PE = VE)$. Experiment 3,4,5 illustrates the case where $(PE < VE)$ on one axis of $PE$ is 1.5 times that of of $VE$, both axis of $PE$ is 1.5 times those of of $VE$, and both axis of $PE$ is 1.85 times those of $VE$. 

\section{Results Analysis}\label{sec:result}
\subsection{Data Collection}
For each designed $(P_{env},V_{env})$ scene pair, we generate a set of three scenes LLM$\&$ARC, ENI++, and ENI++$\&$ARC for user test. 
For each scene, we detected on two aspects: physical collision, and user satisfaction to the room layout. Since ENI++ and ENI++$\&$ARC share identical scenes, user satisfaction of ENI++ was collected whichever comes first in each set of test. User will also start at different locations in identical ENI++ and ENI++$\&$ARC scenes to ensure that subjects fully explore both scenes for collision data collection.

\subsection{Result Analysis: Human Evaluation on Generated Scenes}

\begin{figure}[thpb]
\centering
\centerline{\includegraphics[width=0.5\textwidth]{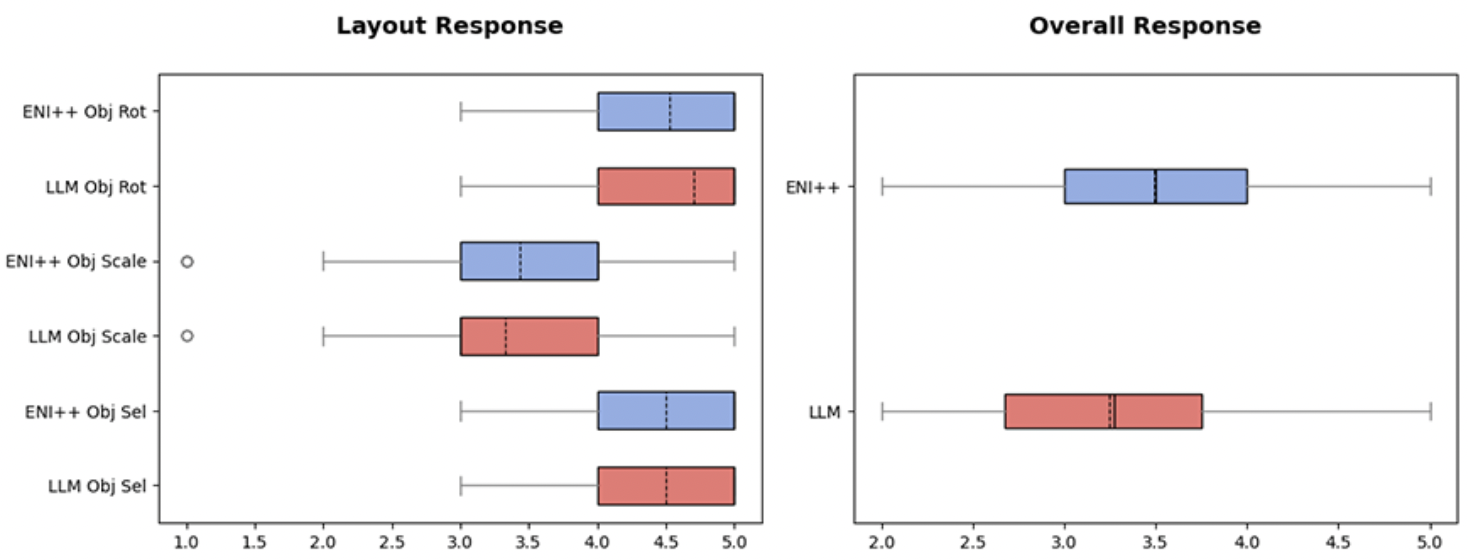}}
\caption{The left figure shows the user responses of objects selection, scaling, and positioning in the both LLM generated scene and scene created under the ENI++ guidance among five different scenarios. The right figure gives the overall users preference to ENI++ or LLM generated room design.}
\label{fig:user_sat}
\vskip -0.1in
\end{figure}

Participants evaluated each room in a scene in four aspects: an overall layout score at the right of Figure \ref{fig:user_sat}, the selection of objects, the scaling of objects and the positioning (placement and rotation) of objects at the left of Figure \ref{fig:user_sat}. They gave a score on the scale of 5 for each aspect. We recorded their responses across all aspects under the scenes LLM and ENI++ or ENI++$\&$ARC. After each set of tests, subjects also chose their preference of layout among them.

Among the four aspects of the evaluation, Figure \ref{fig:user_sat} shows a general favoring of the ENI++ scenes. Figure \ref{fig:user_sat} indicates that there is no clear preference between LLM and ENI++ on their object selection ability. This is because both share the same LLM text prompt and output similar object assets. ENI++ information that prompted in the LLM does not affect the LLM's ability to choose proper 3D models for a room. In object scaling, ENI++ received a slightly higher preference. LLM, on the other hand, indicates a better ability to position objects. ENI++ -directed modification sacrifices a portion of positioning ability to reach a balance between layout design and the compatibility of virtual environments with the physical world. In conclusion, the result demonstrates a comparable user satisfaction for all four scales for ENI++ generation.

In addition to the four evaluated aspects, users choose their preference of layout among two distinct scenes in each set (three scenes, but ENI++ and ENI++$\&$ARC are identical). Among the five sets of scenes, most annotators prefer ENI++ ($50\%$) to LLM$\&$ARC ($37.5\%$), while sometimes no clear inclination ($12.5\%$) was shown.

\begin{table}
\begin{center}
\caption{Entering Virtual Rooms Outside Physical Boundaries}
\label{table:RR}
\begin{tabular}{| c | c | c | c |}
\hline
Methods & LLM$\&$ARC & ENI++ & HCVR (ENI++\&ARC)\\
\hline
Experiment 1& 0.00\% & 0.00\% & 0.00\%\\
\hline
Experiment 2& 0.00\% & 10.40\% & 0.00\%\\ 
\hline
Experiment 3& 3.13\% & 17.20\% & 1.56\%\\
\hline 
Experiment 4& 6.25\% & 34.38\% & 1.56\%\\
\hline 
Experiment 5& 10.00\% & 35.00\% & 6.26\%\\
\hline 
\end{tabular}
\end{center}
\vskip -0.3in
\end{table}

\begin{figure}[h]
\centering
\centerline{\includegraphics[width=0.5\textwidth]{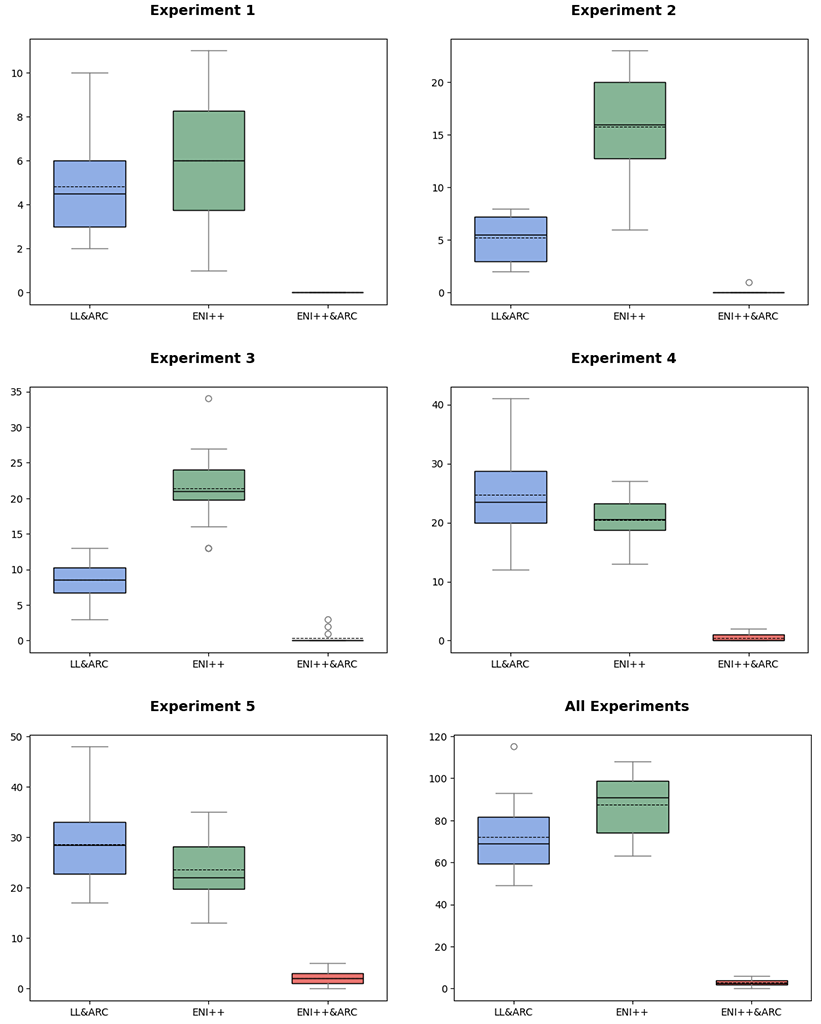}}
\caption{This figure shows the number of detected collisions in the ENI++ -generated scene with the RDW algorithm while walking, the ENI++ scene, and the LLM-generated scene with the RDW algorithm while walking across all five experiments. It visualizes the collision results under different settings of virtual and physical scene pair. The plot illustrates an overall crucially lower collision count than LLM$\&$ARC and ENI++. This highlights the advantages of the ENI++ -directed scene and its compatibility with the alignment-based RDW algorithm in this scenario.} 
\label{fig:collision5}
\end{figure}

\subsection{Result Analysis: Physical Collision Examination}

\subsubsection{Physical Collision Experiments}

We plotted the number of user collisions with physical boundaries and obstacles across five experimental scenes (Figure \ref{fig:collision5}). The results show that \textit{HCVR (ENI++\&ARC)} significantly reduces collisions in all cases. In Experiment 5, where the virtual space is four times larger than the physical space, \textit{HCVR (ENI++\&ARC)} still maintains effective collision management. These results validate that our ENI++ -guided scene generation combined with the RDW algorithm enables near-collision-free navigation in typical scenarios and where LLM\&ARC fails to maintain low collision rates under spatial mismatches.

The first three experiments in Figures \ref{fig:collision5} also illustrates a comparison between ENI++ and LLM$\&$ARC. In Experiments 1, 2 and 3, LLM$\&$ARC demonstrates the ARC advantages in collision avoidance. Comparing with ENI++ as a reference, it's clear to see how ENI++$\&$ARC can fully release the ARC ability in HCVR. Meanwhile, ENI++$\&$ARC obtains a better result given a larger and more complex virtual space as input. Therefore, we conclude that when a more complex virtual scene is given as input, the impact of scene compatibility increases significantly.

\textbf{Experiment 1}
 Figures \ref{fig:collision5} illustrates the collision results when $P_{env}$ is larger than $V_{env}$ in both length and width. Our approach ENI++$\&$ARC was able to achieve 0 collisions in Experiment 1 across all users. 

The robust ANOVA (Analysis of Variance) statistical test revealed a significant difference in the collision count of LLM$\&$ARC, ENI++, ENI++$\&$ARC. The $f = 30.74148$. The $p < .00001$. The result is significant at $p < .05$. As shown in \ref{fig:collision5}, the ENI++$\&$ARC approach achieves a median and mean of 0 where LLM$\&$ARC reaches a median of 4.5 and mean of 4.8125 physical collisions. Without the ARC algorithm, ENI++ reached a median and mean of 6 collisions.

\textbf{Experiment 2}
Figures \ref{fig:collision5} illustrates the collision results when $P_{env}$ shares the same length and width with $V_{env}$. Our approach, ENI++$\&$ARC, successfully achieved zero collisions across users in Experiment 2 except one user got one physical collision during the test.

Our result shows $f = 30.74148$, $p < .00001$. The result is significant at $p < .05$. Figure \ref{fig:collision5} indicates that the ENI++$\&$ARC approach achieves a median of 0 and a mean of 0.0625. LLM$\&$ARC reaches a median of 5.5 and a mean of 5.25 physical collisions. Without the ARC algorithm, ENI++ reached a median of 15.75 and a mean of 16 collisions.

\textbf{Experiment 3}
Figures \ref{fig:collision5} illustrates the collision results when $V_{env}$ is thus half time larger than $P_{env}$. The collision count of our approach is significantly smaller than LLM$\&$ARC to prove the effectiveness of ENI++ -directed designed scene. Meanwhile, it achieves less physical collisions than ENI++ without RDW to prove the strength of ENI++ -directed scene along RDW controller (where we used ARC in our experiment).

A significant difference in the collision count of LLM$\&$ARC, ENI++, ENI++$\&$ARC incurred $f = 147.26238$. The $p < .00001$. The result is significant at $p < .05$. Figure \ref{fig:collision5} plots the advantage of the ENI++$\&$ARC approach which achieves a median of 0 and a mean of 0.375 where LLM$\&$ARC reaches a median and mean of 8.5 physical collisions. Without RDW application during the test test, ENI++ reached a median of 21.4375 and mean of 21 collisions.

\textbf{Experiment 4}
Figures \ref{fig:collision5} visualizes the collision results. when both length and width of $V_{env}$ is half time longer than $P_{env}$. Therefore. $V_{env}$ is 2.25 times of $P_{env}$. The collision count of our approach is significantly smaller than LLM$\&$ARC to prove the effectiveness of ENI++ -directed designed scene. Meanwhile, it achieves less physical collisions than ENI++ without RDW to prove the strength of ENI++ -directed scene along with RDW controller (where we used ARC in our experiment).

Figure \ref{fig:collision5} shows a remarkable result achieved by ENI++$\&$ARC method collected as $f = 95.12786$, $p < .00001$. The result is significant at $p < .05$. The ENI++$\&$ARC approach achieves a median of 0 and a mean of 0.375 where LLM$\&$ARC reaches a median of 23.5 and mean of 24.75 physical collisions. While no redirected controller was applied to ENI++, users reached a median of 20.5 and a mean of 20.4375 collisions.

\textbf{Experiment 5}
Figures \ref{fig:collision5} visualizes the collision results when both length and width of $V_{env}$ is one time larger than $P_{env}$. This leaves $V_{env}$ 4 times of $P_{env}$. The plot illustrates a crucially lower collision count than LLM$\&$ARC and ENI++. This highlights the advantages of the ENI++ -directed scene and its compatibility with the RDW algorithm in this scenario.

While the virtual scene is much larger than the physical space, ENI++$\&$ARC sustains $f = 98.75979$, p $<$ .00001. The result is significant at $p < .05$. As shown in Figure \ref{fig:collision5}, the ENI++$\&$ARC approach achieves a median of 2 and mean of 2.125 where LLM$\&$ARC reaches a median of 28.5 and mean of 28.6875 physical collisions. Without the ARC algorithm, ENI++ reached a median of 22 and mean of 23.6875 collisions. 

\subsubsection{Generated Scene Compatibility Analysis} 

 Among five experiments described in last section, we run ENI++ as a quantitative metrics on generated virtual scenes to evaluate each physical space compatibility. Table \ref{table:ENI++_Score} illustrates average ENI++ score comparison between LLM-directed and ENI++-directed scenes (where smaller number represents a higher compatibility). We demonstrate that our architecture is able to provide a compatible virtual scene and therefore more likely to redirect users on a collision-free path while free walking.

\begin{table}
\begin{center}
\caption{ENI++ Score Across LLM-directed and ENI++-directed scenes}
\label{table:ENI++_Score}
\begin{tabular}{| c | c | c | c |}
\hline
Experiments & Floor Plan & LLM-directed & ENI++-directed \\
\hline
Experiment 1& 0 & 0 & 0 \\
\hline
Experiment 2& 20.81 & 13.54 & 1.45 \\ 
\hline
Experiment 3& 528.74 & 376.49 & 168.07 \\
\hline 
Experiment 4& 1187.63 & 852.72 & 352.54 \\
\hline 
Experiment 5& 2624.06 & 1763.35 & 765.89 \\
\hline 
\end{tabular}
\end{center}
\vskip -0.3in
\end{table}

\subsubsection{Unreachable Rooms Analysis} \label{sec:unreachable}

Virtual scenes is larger than physical scenes in some experiments, and not all rooms are accessible to subjects in that case. This means users have to walk outside of the physical designed boundary to explore some of the virtual rooms. Due to maintaining user a more immersive walking experience for layout rating, we do not reset the virtual scene when user walk out of the physical boundary. Walking out of physical boundary was counted as normal collisions and unreachable rooms can count as a supplementary for the above collision data. This means when user walking out of physical boundary while exploring a room, it is marked as unreachable. The ARC technique terminates until the user enter back to the designed physical space. We counted the unreachable room for each experiment trail and plotted the average possibility that users have an unreachable room situation in Table \ref{table:RR}.

ENI++$\&$ARC shows a notable result among all experiments. Without ARC applied during the experiments, the users would easily walk out of the physical area during the scene exploration. ENI++ -guided scene generation also prevents users from walking out of the physical room boundary while the same RDW controller is shared in LLM$\&$ARC and ENI++$\&$ARC.

\section{Conclusion, Limitations, and Future Works}
We introduce HCVR, a novel interface that leverages large language models to generate VR scenes tailored to diverse physical environments, enabling VR users to explore freely via redirected walking. To quantify scene–environment alignment, we propose the ENI++ score, which evaluates the compatibility of each virtual and physical area pair. By comparing LLM-driven layout, ENI++ -directed layout, HCVR achieves an optimal balance between coherent design and seamless integration of virtual and physical spaces. This paper also quantitatively illustrate how the ability of HCVR pipeline on unleashing the redirected walking ability of ARC and direct user onto collision-free path through virtual scene design.

\subsection{Number of Collision VS Reset Evaluation}
Most RDW algorithms implement a reset component \cite{williams2021arc, nguyen2018discrete, zhang2022one} to assess their methods where each reset that stop users (through symbols, distractors, and etc.) whenever a physical collision with physical boundaries is recorded. In this paper, we adopted an more intuitive approach to collect collision data to avoid influencing user satisfaction responses. Users' perception of natural walking in virtual environments can positively impact their interaction and reaction to a scene by enhancing the sense of embodiment (SoE) \cite{spangenberger2024embodying,calogiuri2018experiencing}.

During the experiments, once the user starts colliding with a physical obstacle, our method disables the ARC algorithm until the collision finishes and accumulates one to the collision data. The collision data is also accumulated again once the user enter back to the physical area. Since our testing scene is larger than the virtual space, the distance of users' free walking without ARC can also be tolerated.

\subsection{Scaling of Objects}
In the result demonstration phase, the user satisfaction result in Figure \ref{fig:user_sat} records a low score on the overall aspect of the object scaling. Both LLM suggested output and OOB procedure can scale objects in our pipeline. During the experiment, the users also reflected on the issue of the disproportion of the height of the objects. This might be caused by the rescaling procedure during the OOB checker and ENI++ -directed modification. The object height re-scaling is calculated from the average of its width and length re-scaling. The unrealistic scaled objects can also be caused by the limited spatial reasoning ability of the LLM model we used (gpt-4o \cite{achiam2023gpt}). Some LLMs specifically developed for 3D tasks \cite{sun2024layoutvlm, spatiallm} can be utilized in future to mitigate the hallucination. A preferred way can also be to utilize or train a 3D-sensitive and object understanding model to scale its height based on its new width and length. This can be a new research problem that can strength HCVR with better user experience.

\subsection{Dynamic Obstacles}
Although HCVR's results were promising, it is also important to acknowledge the limitations of this study. Our experiment does not include any dynamic obstacles during the user test. This is because a physical environment with moving obstacles is not able to create a static ENI++ map, which is essential for the room generation and later procedure. Meanwhile, it is hard to compute the compatibility for moving obstacles in physical environment. We believe our model can be well-combined with saccade behavior redirection \cite{sun2018towards} method to handle dynamic obstacles during the test. We would like to extend our approach to handle both static and dynamic physical environment. Our approach will focus on combining the real-time saccade redirection and collecting data on eye movement along with the human response to object movement.

\clearpage

\bibliographystyle{abbrv-doi-hyperref}
\bibliography{template}
\end{document}